\documentclass[twocolumn,english,prl,showpacs,aps]{revtex4-1}

\usepackage[T1]{fontenc}
\usepackage[latin9]{inputenc}
\usepackage{amsmath}
\usepackage{amssymb}
\usepackage{graphicx}

\usepackage{babel}

\makeatletter
 
 \@ifundefined{textcolor}{}
 {%
   \definecolor{BLACK}{gray}{0}
   \definecolor{WHITE}{gray}{1}
   \definecolor{RED}{rgb}{1,0,0}
   \definecolor{GREEN}{rgb}{0,1,0}
   \definecolor{BLUE}{rgb}{0,0,1}
   \definecolor{CYAN}{cmyk}{1,0,0,0}
   \definecolor{MAGENTA}{cmyk}{0,1,0,0}
   \definecolor{YELLOW}{cmyk}{0,0,1,0}
 } 

\@ifundefined{textcolor}{}{%
 \definecolor{BLACK}{gray}{0}
 \definecolor{WHITE}{gray}{1}
 \definecolor{RED}{rgb}{1,0,0}
 \definecolor{GREEN}{rgb}{0,1,0}
 \definecolor{BLUE}{rgb}{0,0,1}
 \definecolor{CYAN}{cmyk}{1,0,0,0}
 \definecolor{MAGENTA}{cmyk}{0,1,0,0}
 \definecolor{YELLOW}{cmyk}{0,0,1,0}
 }

\@ifundefined{definecolor}{\@ifundefined{definecolor}
 {\@ifundefined{definecolor}
 {\@ifundefined{definecolor}
 {\@ifundefined{definecolor}
 {\@ifundefined{definecolor}
 {\usepackage{color}}{}
}{}
}{}
}{}
}{}
}{}
\makeatother

\begin{document}

\title{Proposal for demonstrating the Hong-Ou-Mandel effect with matter waves}

\author{R.~J.~Lewis-Swan and K.~V.~Kheruntsyan$^{*}$}

\affiliation{The University of Queensland, School of Mathematics and Physics, Brisbane, Queensland 4072, Australia.\\
$^{(*)}$Correspondence and requests for materials should be addressed to K.V.K. (email <karen.kheruntsyan@uq.edu.au>).}

\date{\today }
\begin{abstract}
The Hong-Ou-Mandel (HOM) effect is a striking demonstration of destructive quantum interference between pairs of indistinguishable bosons, realised so far only with massless photons. Here we propose an experiment which can realise this effect in the matter-wave regime using pair-correlated atoms produced via a collision of two Bose-Einstein condensates and subjected to two laser induced Bragg pulses. We formulate a novel measurement protocol appropriate for the multimode matter-wave field, which---unlike the typical two-mode optical case---bypasses the need for repeated measurements under different displacement settings of the beam-splitter, thus dramatically reducing the number of experimental runs required to map out the interference visibility. The protocol can be utilised in related matter-wave schemes; here we focus on condensate collisions and by simulating the entire experiment we predict a HOM-dip visibility of $\sim69$\%. By being larger than 50\%, such a visibility highlights strong quantum correlations between the atoms and paves the way for a possible demonstration of a Bell inequality violation with massive particles in a related Rarity-Tapster setup.
\end{abstract}


\maketitle

~\\
Since its first demonstration, the Hong-Ou-Mandel (HOM) effect \cite{HOM-87} has become
a textbook example of quantum mechanical 
two-particle interference using pairs of indistinguishable photons. When two such photons enter a $50$:$50$ beam
splitter, with one photon in each input port, they both preferentially
exit from the same output port, even though each photon individually
had a $50$:$50$ chance of exiting through either output port. The HOM effect was 
first demonstrated
using optical parametric down-conversion \cite{HOM-87}; the same setup, but
with an addition of linear polarisers, was subsequently used to demonstrate
a violation of a Bell inequality \cite{Ou-Mandel-88} which is of fundamental importance to validating some of the foundational principles of quantum mechanics such as quantum nonlocality and long-distance entanglement.

The HOM effect is 
a result of destructive 
quantum 
interference in a (bosonic) twin-photon state, which
leads to a characteristic dip 
in the photon coincidence counts at
two photodetectors placed at the output ports of a beam splitter. 
The destructive 
interference occurs between
two \textit{indistinguishable} paths 
corresponding to the photons being both reflected from, or both transmitted through, the beam splitter.
Apart from being of fundamental importance to quantum physics, the HOM effect 
underlies the basic entangling mechanism in linear 
optical quantum computing \cite{Linear_quantum_computation}, in which a twin-photon state $|1,1\rangle$ is converted into a quantum 
superposition 
$\frac{1}{\sqrt{2}}(|2,0\rangle - |0,2\rangle)$ --- the simplest example of the elusive `NOON' state \cite{Dowling-NOON}.
Whereas the HOM effect with (massless) photons has been extensively studied in quantum optics (see \cite{Duan-Monroe-2010,HOM-microwave} and references therein), two-particle quantum interference 
with massive particles remains largely unexplored. A matter-wave 
demonstration of the HOM effect 
would be a major advance in experimental quantum physics, enabling an expansion of foundational tests of quantum mechanics 
into previously unexplored regimes.

\begin{figure}
\includegraphics[width=8.1cm]{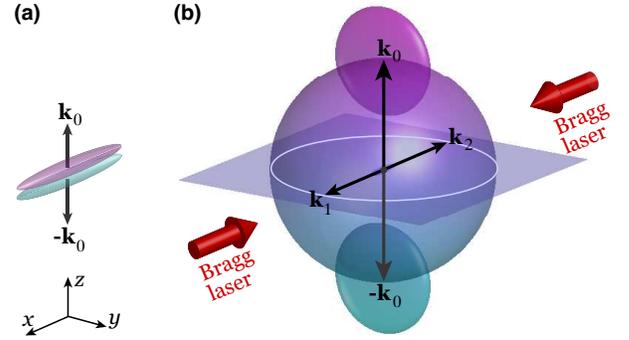} 
\caption{\textbf{Schematic diagram of the collision geometry.} (a) Position space picture of two elongated (along the $x$-axis) Bose-Einstein condensates counter propagating along $z$ with equal but opposite momenta $\pm\mathbf{k}_0$. (b) Momentum space distribution of the atomic cloud showing the (disk shaped) condensates on the north and 
south poles of the spherical halo of scattered atoms (see text for further details). }
\label{fig:scheme} 
\end{figure}

\begin{figure*}[tpp]
\includegraphics[width=17.9cm]{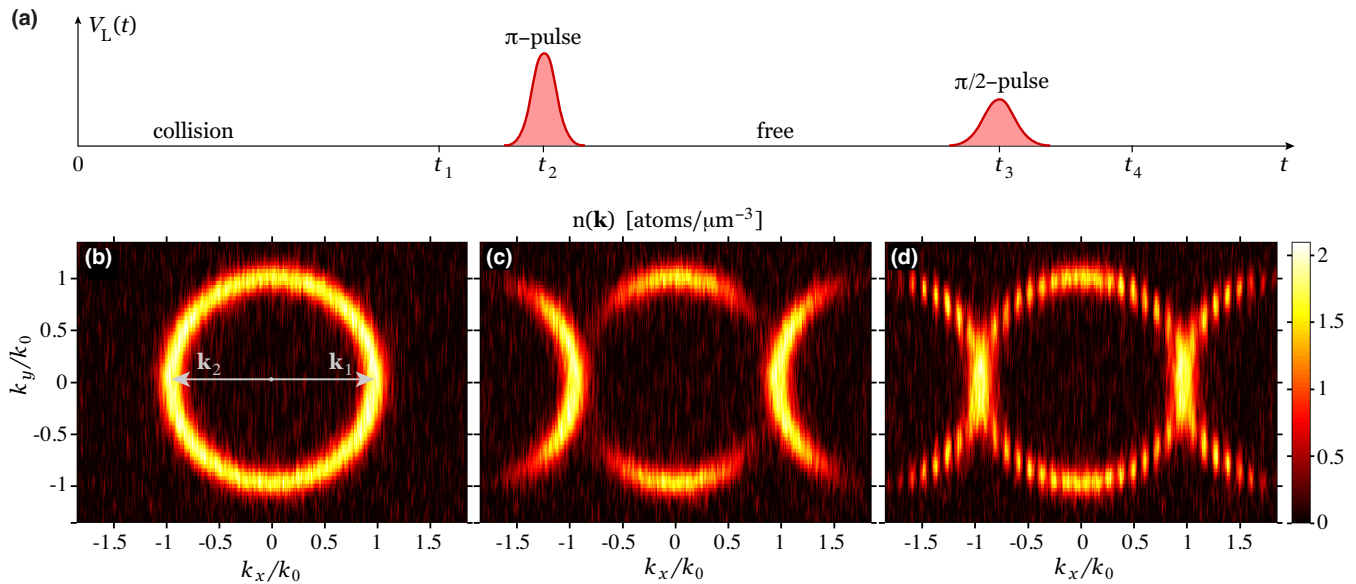} 
\caption{\textbf{Timeline of the proposed experiment and snapshots of the collisional halo from the numerical simulations.} In panel (a), time $t=0$ corresponds to the coherent splitting of the source condensate that sets up the collision; $V_{L}(t)$ denotes the depth of the lattice potential formed by the Bragg 
lasers, with the first hump indicating the mirror ($\pi$) pulse, while the second hump -- the beam-splitter ($\pi/2$) pulse (the initial source-splitting pulse is not shown for clarity). 
Panels (b)-(d) show the results of numerical simulations of the momentum-space density 
distribution $n(\mathbf{k})$ of scattered atoms
on the equatorial plane of the halo: (b) shows the density distribution after the collision, at $t_1=65$ $\mu$s; (c) -- after the $\pi$-pulse,
centred at $t_2=75$ $\mu$s and having a duration of $\tau_{\pi}=2.5$ $\mu$s (rms width of Gaussian envelope); and (d) -- after the final
$\pi/2$ pulse, with $\Delta t_{\mathrm{free}}=t_{3}-t_{2}=85$ $\mu$s and $\tau_{\pi/2}=2.5$ $\mu$s 
(see Methods for further details; the durations shown on the time axis are not to scale).
The momentum axes $k_{x,y}$ in panels (b)-(d)
are normalised to the collision momentum $k_{0}\equiv|\mathbf{k_{0}}|$
(in wave-number units), which in our simulations was $k_{0}=4.7\times10^{6}$
m$^{-1}$. The simulations were carried
out for an initial BEC containing a total of $N=4.7\times10^{4}$ atoms
of metastable helium ($^{4}$He$^{*}$), prepared in a harmonic trap of frequencies
$(\omega_{x},\omega_{y},\omega_{z})/2\pi=(64,1150,1150)$ Hz, and
colliding with the scattering length of $a=5.3$ nm; all these parameters
are very close to those realised in recent experiments \cite{Jaskula-10,Krachmalnikoff:2010}. 
\label{fig:Collision-halo-pics} 
}
\end{figure*}

Here we propose an experiment which can realise 
the HOM effect with matter waves using a collision of two atomic Bose-Einstein 
condensates (BECs) (as in Refs.~\cite{Perrin:2007,Perrin-08,Krachmalnikoff:2010,Jaskula-10,Kheruntsyan:2012}) and a pair of laser-induced Bragg pulses. The HOM interferometer uses pair-correlated atoms from the scattering halo that is generated during the collision through the process of spontaneous four-wave mixing.
The pair-correlated atoms are mixed using two separate Bragg pulses \cite{Kozuma-99,meystre-atom-optics} that realize an atom-optics mirror and beam-splitter elements---in analogy with the use of twin-photons from parametric down-conversion in the optical HOM interferometer scheme. The HOM effect is quantified via the measurement of a set of atom-atom pair correlation functions  between the output ports of the interferometer. Using stochastic quantum simulations of the collisional dynamics and the application of Bragg pulses, 
we predict a HOM-dip visibility of $\sim\!69$\% for realistic experimental parameters. 
A visibility larger than $\sim\!50$\% is 
indicative of stronger than classical correlations between the atoms in the scattering halo \cite{Vienna-twins,Jaskula-10,Hannover-twins,Oberthaler-quadratures,Kheruntsyan:2012}, which in turn renders our system as a suitable platform for demonstrating a Bell's inequality violation with matter waves using a closely related Rarity-Tapster scheme \cite{Rarity:90}. \\
\\
\textbf{\large{Results}}\\
\textbf{Setup.} The schematic diagram of the proposed experiment is shown in Fig. \ref{fig:scheme}.
A highly elongated (along the $x$-axis) BEC is initially split into two equal and 
counterpropagating halves 
traveling with momenta $\pm\mathbf{k}_{0}$ along $z$ in
the centre-of-mass frame. Constituent atoms undergo binary elastic
collisions which produce a nearly spherical $s$-wave scattering
halo of radius {$k_{r}\simeq 0.95|\mathbf{k}_{0}|$} \cite{Krachmalnikoff:2010} in
momentum space due to energy and momentum conservation. The elongated
condensates have a disk shaped density distribution in momentum space, shown in Fig.~\ref{fig:scheme}~(b) 
on the north and south poles of the halo. After the end of the
collision (which in this geometry corresponds to complete spatial
separation of the condensates in position space) we apply two counterpropagating
lasers along the $x$-axis whose intensity and frequency are tuned
to act as a resonant Bragg $\pi$-pulse with respect to two diametrically opposing
momentum modes, $\mathbf{k}_{1}$
and $\mathbf{k}_{2}=-\mathbf{k}_{1}$, situated on the equatorial plane of the halo and satisfying $|\mathbf{k}_{1,2}|\!=\!k_{r}$.

Previous
experiments and theoretical work \cite{Perrin:2007,Jaskula-10,Kheruntsyan:2012,Gardiner:06,Savage:06,Deuar:07,Moelmer:2008,Perrin-08,Ogren-09,Deuar-11}
have shown the existence of strong atom-atom correlation between such {diametrically opposite} modes, similar to the correlation between twin-photons in parametric down-conversion. 
Applying the Bragg 
$\pi$-pulse {to the collisional halo} replicates an optical mirror and reverses the trajectories
of the scattered atoms with momenta $\mathbf{k}_{1}$ and $\mathbf{k}_{2}$,
and a finite region around them. {We assume that the pulse is tuned to operate in the so-called Bragg regime of the Kapitza-Dirac effect \cite{meystre-atom-optics,Batelaan-00} (diffraction of a matter-wave from a standing light field), 
corresponding to conditions in which second- and higher-order diffractions are suppressed.}
The system is then allowed to propagate
freely for a duration so that the targeted atomic wave-packets regain spatial overlap
in position space.
We then apply a second Bragg pulse -- a $\pi/2$-pulse -- to replicate
an optical $50$:$50$ beam-splitter, which is again targeted to couple  
$\mathbf{k}_{1}$ and $\mathbf{k}_{2}$, thus realising the HOM interferometer.

\begin{figure*}[tpp]
\includegraphics[width=17.7cm]{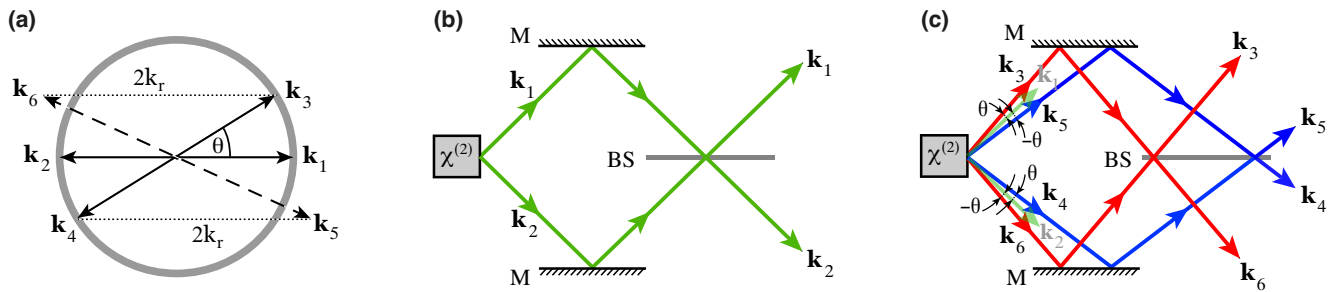} 
\caption{\textbf{Schematic of matter-wave momentum modes affected by the Bragg pulses and the topologically equivalent optical setups.}
In panel (a), the diametrically opposite vectors $\mathbf{k}_{1}$ and $\mathbf{k}_{2}=-\mathbf{k}_{1}$
show the targeted momentum modes on the scattering halo; their amplitude is given by the
halo peak radius, $k_{r}=|\mathbf{k}_{1}|=|\mathbf{k}_{2}|$, which is equal to $k_r=0.95|\mathbf{k}_{0}|$ in this part of the halo
\cite{Krachmalnikoff:2010}. Also shown are the to-be-measured momentum components $\mathbf{k}_3$ and $\mathbf{k}_4$ 
corresponding to a rotation by $\theta$ away from the targeted modes, 
which couple, respectively, to $\mathbf{k}_6 = \mathbf{k}_{3} - 2\mathbf{k}_1$ and $\mathbf{k}_5 = \mathbf{k}_{4} + 2\mathbf{k}_1$ by the same Bragg pulses. 
Panels (b) and (c) show equivalent optical schemes which utilise a $\chi^{(2)}$ nonlinear crystal that is optically pumped to produce twin-photons via parametric 
 down-conversion. In (b) we depict the archetypal optical HOM setup which corresponds to the case of $\theta = 0$ in the matter-wave scheme. 
 A twin-photon state in modes $\mathbf{k}_1$ and $\mathbf{k}_2$ is first selected from a 
 broadband source, then mixed at the beam-splitter (BS) after reflection from the mirror (M), and photon coincidence counts are 
 measured between the two symmetric output ports of the interferometer. 
 In (c) we depict the optical 
 setup which is equivalent to $\theta > 0$ in the mater-wave proposal. Two twin-photon states in ($\mathbf{k}_3,\mathbf{k}_4$) and in ($\mathbf{k}_5,\mathbf{k}_6$)
 are selected from  the broadband source; the asymmetry of the pairs about the optical axis of the interferometer means 
 that the correlated photons from the respective pairs will arrive at the beam-splitter at spatially separate locations and will mix with photons from the other pair, which introduces distinguishability between the paths through the interferometer.
\label{fig:modes-and-optical-scheme} 
}
\end{figure*}

The timeline of the proposed experiment is illustrated in figure 
\ref{fig:Collision-halo-pics}~(a), whereas the results of numerical simulations (see Methods)
of the collision dynamics and the application of Bragg pulses are shown in figures \ref{fig:Collision-halo-pics}~(b)-(d):
(b) shows the 
equatorial slice of the momentum-space density distribution $n(\mathbf{k},t)$ of the 
scattering halo at the end of collision; (c) and
(d) show the halo density after the application of the $\pi$ and $\pi/2$ pulses, respectively. The `banana' shaped regions in (c) correspond to `kicked'
populations between the targeted momenta around $\mathbf{k}_{1}$
and $\mathbf{k}_{2}$ in the original scattering halo, while (d) shows the density distribution after mixing. 
The density modulation in (c) is simply the result of interference between the residual and transferred atomic populations after the $\pi$-pulse 
upon their recombination on the beamsplitter.
The residual population is due to the fact that the pairs of off-resonance modes in these parts of the halo (which are coupled by the same Bragg pulses as they share the 
same momentum difference $2k_r$ as the resonant modes $\mathbf{k}_{1}$ and $\mathbf{k}_{2}$) 
no longer satisfy the perfect Bragg resonance condition and therefore the population transfer during the $\pi$-pulse is not $100$\% efficient. 
As these components have unequal absolute momenta, their amplitudes accumulate a 
nonzero relative phase due to phase dispersion during the free propagation. The accrued relative phase results in interference fringes upon the recombination on the beamsplitter, with an approximate period of  {$\Delta k\simeq \pi m/(\hbar k_r \Delta t_{\mathrm{free}}) \simeq 0.1 |\mathbf{k}_0|$}.

Due to the indistinguishability of the paths of the Bragg-resonant modes $\mathbf{k}_{1}$ and $\mathbf{k}_{2}$ through the beam-splitter and the resulting destructive quantum interference, a measurement of 
coincidence counts between the atomic populations in these modes will 
reveal a suppression compared to the background level. 
To reveal the full structure of the HOM dip, including the background level where no quantum interference occurs, we must introduce 
path distinguishability between the $\mathbf{k}_{1}$ and $\mathbf{k}_{2}$ modes. One way to achieve this, which would be in a direct analogy with shifting the beam splitter in the optical HOM 
scheme, is to change the Bragg-pulse resonance condition from the ($\mathbf{k}_1$, $\mathbf{k}_2$) pair to ($\mathbf{k}_1$, $\mathbf{k}_2+\hat{\mathbf{e}}_x\delta k$), where $\hat{\mathbf{e}}_x$ 
is the unit vector in the $x$-direction. The approach to the background coincidence 
rate between the populations in the $\mathbf{k}_{1}$ and $\mathbf{k}_{2}$ modes would then correspond to 
performing the same experiment for increasingly large displacements $\delta k$. 
Taking into account that acquiring statistically significant results for each $\delta k$ requires 
repeated runs of the experiment (typically thousands), this measurement protocol could potentially pose a significant practical challenge due to the very large total number of experimental 
runs required.
\\
\\
\textbf{Proposed measurement protocol.}
To overcome this challenge, we propose an alternative measurement protocol 
which can reveal the full structure of the HOM dip from just one Bragg-resonance condition, 
requiring only one set of experimental runs. 
The protocol takes advantage of the broadband, multimode nature of the scattering halo and the fact that the original Bragg pulse couples not only the 
targeted momentum modes $\mathbf{k}_{1}$ and $\mathbf{k}_{2}$, but also many other pairs of modes which follow distinguishable paths through the beam-splitter. One such 
pair, $\mathbf{k}_{3} = (k_x,k_y,k_z) = k_r(\mathrm{cos}(\theta),\mathrm{sin}(\theta),0)$ and $\mathbf{k}_{4} = -\mathbf{k}_{3}$, located on the halo peak, is shown in 
Fig. \ref{fig:modes-and-optical-scheme}~(a) and corresponds to a rotation by angle $\theta$ away from $\mathbf{k}_{1}$ and $\mathbf{k}_{2}$. 
The modes $\mathbf{k}_{3}$ and $\mathbf{k}_{4}$ are equivalent to the original pair in the sense of their quantum statistical properties and therefore, these modes can be used for the 
measurement of the background level of coincidence counts, instead of physically altering the paths of the $\mathbf{k}_{1}$ and $\mathbf{k}_{2}$ modes. The angle $\theta$ now serves the role of the 
`displacement' parameter that scans through the shape of the HOM dip. A topologically equivalent optical scheme {is shown in Figs.~\ref{fig:modes-and-optical-scheme}(b)-(c),} which is in turn similar to the one analysed in 
Ref. \cite{Rarity-89} using a broadband source of angle-separated 
pair-photons and directionally asymmetric apertures.

\begin{figure}
\includegraphics[width=8cm]{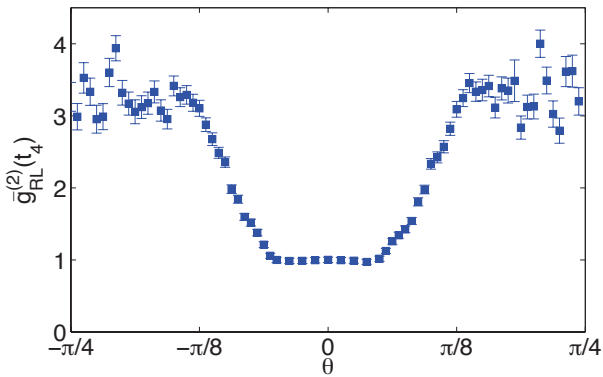}
\caption{\textbf{Hong-Ou-Mandel dip.}
Normalised atom-atom correlation function $\overline{g}^{(2)}_{\mathrm{RL}}(t_4)$ between the two arms of the interferometer, characterising the HOM effect as a function of the path-distinguishability 
angle $\theta$. Error bars denote sampling error from $\sim \!30,000$ stochastic simulations (see Methods). The atom counting bins are rectangular boxes with sides $\delta k_{x}\!=\!0.01k_{0}$ and 
$\delta k_{y,z}\!=\!0.19k_{0}$ which approximate the widths of the momentum distribution of the initial trapped BEC. 
}
\label{fig:HOM-dip}
\end{figure}

In the proposed protocol, detection {(after the final Bragg pulse) of atom coincidences at the pair of originally correlated momenta $\mathbf{k}_{3}$ and $\mathbf{k}_{4}$ corresponds to both paths being separately \textit{reflected} on 
the beamsplitter (see Fig. \ref{fig:modes-and-optical-scheme}(c)). Apart from this outcome, we need to take into account the coincidences between the respective Bragg-partner momenta, $\mathbf{k}_{6}$ and $\mathbf{k}_{5}$ (separated, respectively, 
from $\mathbf{k}_{3}$ and $\mathbf{k}_{4}$ by the same difference $2k_r$ as $\mathbf{k}_{1}$ from $\mathbf{k}_{2}$). Coincidences at $\mathbf{k}_{6}$ and $\mathbf{k}_{5}$ correspond to atoms of the originally correlated momenta $\mathbf{k}_{3}$ and $\mathbf{k}_{4}$ being both \textit{transmitted} through the beam splitter (see Fig. \ref{fig:modes-and-optical-scheme}(c)). Finally, in order to take into account all possible channels contributing to coincidence counts between the two arms of the interferometer, 
we need to measure coincidences between $\mathbf{k}_{3}$ and $\mathbf{k}_{6}$, as well as between $\mathbf{k}_{4}$ and $\mathbf{k}_{5}$. This ensures that the total detected flux at the output ports of the beam splitter matches the total input flux.
In addition to this, we normalise the bare coincidence counts to the product of single-detector count rates, \textit{i.e.}, the product of the average number of atoms in the two output arms of the interferometer. 
We use the normalised correlation function as the total population in the four relevant modes varies as the angle $\theta$ is increased, implying that the raw coincidence rates 
are not a suitable quantity to compare at different angles.}
\\
\\
\textbf{Hong-Ou-Mandel effect and visibility of the dip.}
With this measurement protocol in mind, we quantify the HOM effect using the normalised second-order correlation
function $\overline{g}^{(2)}_{\mathrm{RL}}(t)\! = \!\langle:\!\!\hat{N}_{\mathrm{R}}(t)\hat{N}_{\mathrm{L}}(t)\!\!:\rangle/ \langle \hat{N}_{\mathrm{R}}(t) \rangle \langle \hat{N}_{\mathrm{L}}(t) \rangle$ 
after the $\pi/2$-pulse concludes at $t \!= \!t_4$. Here, $\langle \hat{N}_{\mathrm{R}} \rangle \!\equiv \!\langle \hat{N}_3 \rangle \!+ \!\langle \hat{N}_5 \rangle$ 
and $\langle \hat{N}_{\mathrm{L}} \rangle \!\!\equiv \!\!\langle \hat{N}_4 \rangle \!+ \!\langle \hat{N}_6 \rangle$ correspond to the number of atoms detected, respectively, on 
the two (right and left) output ports of the beam splitter, with the detection bins centred around the four momenta of interest $\mathbf{k}_i$ ($i \!= \!3,4,5$, and $6$), for any given 
angle $\theta$ [see Fig. \ref{fig:Collision-halo-pics}(e)]. More specifically, $\hat{N}_{i}\left(t\right)\!=\!\intop_{\mathcal{V}(\mathbf{k}_{i})}\!d^{3}\mathbf{k\,}\hat{n}\left(\mathbf{k},t\right)$ 
is the atom number operator in the integration volume $\mathcal{V}(\mathbf{k}_{i}$) centred around $\mathbf{k}_i$, where $\hat{n}(\mathbf{k},t)=\hat{a}^{\dagger}(\mathbf{k},t)\hat{a}(\mathbf{k},t)$
is the momentum-space density operator, with $\hat{a}^{\dagger}(\mathbf{k},t)$ and $\hat{a}(\mathbf{k},t)$ the corresponding creation 
and annihilation operators (the Fourier components of the field operators $\hat{\delta}^{\dag}(\mathbf{r},t)$ and $\hat{\delta}(\mathbf{r},t)$, see Methods). The double-colon notation 
in $\langle:\!\!\hat{N}_{\mathrm{R}}(t)\hat{N}_{\mathrm{L}}(t)\!\!:\rangle$ indicates normal ordering of the creation and annihilation operators.

The integrated form of the second-order correlation correlation function, which quantifies the correlations in terms of atom number coincidences in detection bins of certain size rather than in terms of local density-density correlations, accounts for limitations in the experimental detector resolution, in addition to improving the signal-to-noise ratio which is typically low due to the relatively low density of the scattering halo; in typical condensate collision experiments and in our simulations, the low density translates to a typical halo-mode occupation of $\sim\!0.1$. We choose $\mathcal{V}(\mathbf{k}_i)$ to be a rectangular box with dimensions corresponding to the rms width of the 
initial momentum distribution of the trapped condensate, which is a reasonable
approximation to the mode (or coherence) volume in the scattering halo \cite{Perrin-08, Ogren-09}.

The second-order correlation function $\overline{g}^{(2)}_{\mathrm{RL}}(t_4)$, quantifying the HOM effect as a function of the path-distinguishability angle $\theta$, is shown in 
Fig.~\ref{fig:HOM-dip}. For $\theta\!=\!0$, where $\mathbf{k}_{3(4)}\!=\!\mathbf{k}_{1(2)}$, we observe maximum suppression of coincidence counts relative to the background level 
due to the indistinguishability of the paths. 
As we increase $|\theta| > 0$, we no longer mix $\mathbf{k}_3$ and $\mathbf{k}_4$ as a pair and their paths through the beam-splitter become 
distinguishable; the path interference is lost, and we observe an increase in the magnitude of the correlation function to the background level.
We quantify the visibility of the HOM dip via $V \!=\! 1 \!- \!\min\{\overline{g}^{(2)}_{\mathrm{RL}}(t_4)\}/\max\{\overline{g}^{(2)}_{\mathrm{RL}}(t_4)\}$, where $\min\{\overline{g}^{(2)}_{\mathrm{RL}}(t_4)\}$ occurs 
for $\theta\!=\!0$ and $\mathrm{max}\{\overline{g}^{(2)}_{\mathrm{RL}}(t_4)\}$ for sufficiently large $\theta$ such that momenta $\mathbf{k}_{5,6}$ lie outside the scattering halo. Due to the oscillatory nature of the wings 
(see below) we take $\mathrm{max}\{\overline{g}^{(2)}_{\mathrm{RL}}(t_4)\}$ to correspond to the mean of $\overline{g}^{(2)}_{\mathrm{RL}}(t_4)$ for $\theta \gtrsim \pi/8$. Using this definition we measure a visibility of  
$V\!\simeq\! 0.69\pm0.08$, where the uncertainty of $\pm0.08$ corresponds to taking into account the full fluctuations of $g^{(2)}_{\mathrm{RL}}(t_4)$ about the mean in the wings rather than fitting the oscillations 
(see Supplementary Fig. 1 and Supplementary Notes 1 and 2). The visibility larger than $0.5$ is consistent with the nonclassical effect of violation of Cauchy-Schwarz inequality 
with matter waves \cite{Ogren-09}, observed recently in condensate collision experiments \cite{Kheruntsyan:2012}. The exact relationship between the visibility and the Cauchy-Schwarz inequality is discussed further 
in Supplementary Note 3, as are simple (approximate) analytic estimates of the magnitude of the HOM dip visibility (Supplementary Notes 1-5 and Supplementary Fig. 2).
\\
\\
\textbf{Width of the HOM dip.} The broadband, multimode nature of the scattering halo implies that the range of the path-length difference 
over which the HOM effect can be observed is determined by the spectral width of the density profile of the scattering halo. 
Therefore the width of the HOM dip is related to the width of the halo density.
This is similar to the situation analysed in Ref. \cite{Rarity-89} using pair-photons from a broadband parametric down-converter.
The angular width of the HOM dip extracted from Fig. \ref{fig:HOM-dip} is approximately $w_{\mathrm{HOM}}\!\simeq \!0.61$ radians, which is indeed close to the width (full width at half maximum) of 
the scattering halo in the relevant direction, $w_{\mathrm{halo}}\!\simeq \!0.69$ radians (see also Supplementary Note 2 for simple analytic estimates). The same multimode nature of the 
scattering halo contributes to the oscillatory behaviour in the wings of the HOM dip profile: here we mix halo modes with unequal absolute momenta and the resulting phase dispersion from free-propagation 
leads to oscillations similar to those observed with two-color photons \cite{Rarity-89}.
\\
\\
\textbf{Comparison with optical parametric down-conversion.}
We emphasise that the input state in our matter-wave HOM interferometer is subtly different from the idealised twin-Fock state $|1,1\rangle$ used in the simplest analytic descriptions of the optical HOM effect. This idealised 
state stems from treating the process of spontaneous optical parametric down-conversion (SPDC) in the weak-gain regime. We illustrate this approximation by considering a two-mode toy model of the process, which in 
the undepleted pump approximation is described by the Hamiltonian $\hat{H} = \hbar{g}(\hat{a}_1^{\dagger}\hat{a}_2^{\dagger} + h.c.)$ that produces perfectly correlated photons in the $\hat{a}_1$ and $\hat{a}_2$ modes, 
where $g>0$ is a gain coefficient related to the quadratic nonlinearity of the medium and the amplitude of the coherent pump beam. (In the context of condensate collisions, the coupling $g$ corresponds to $g=U\rho_0(0)/\hbar$ at the same level of `undepleted pump' approximation \cite{Perrin-08,Ogren-09}; see Methods for the definitions of $U$ and $\rho_0$.) The full output state of the SPDC process in the Schr\"{o}dinger picture is given by 
$|\psi\rangle = \sqrt{1-\alpha^2} \sum_{n=0}^{\infty} \alpha^n |n,n\rangle$, where $\alpha = \mathrm{tanh}(g{t})$ and $t$ is the interaction time \cite{Braunstein-review}. In the weak-gain regime, corresponding to
$\alpha\simeq g{t} \ll 1$, this state is well approximated by $|\psi\rangle \propto |0,0\rangle + \alpha|1,1\rangle$, \textit{i.e.}, by truncating the expansion of $|\psi\rangle$ and neglecting the contribution of the $|2,2\rangle$ and higher-$n$ components.
This regime corresponds to mode populations being much smaller than one, $\langle \hat{n} \rangle=\langle \hat{a}_{1(2)}^{\dagger}\hat{a}^{~}_{1(2)} \rangle =\sinh^2(g t)\simeq (g t)^2\simeq \alpha^2 \ll1$. The truncated state itself is qualitatively
identical to the idealised state $|1,1\rangle$ as an input to the HOM interferometer: both result in a HOM dip minimum of 
$\bar{g}^{(2)}_{\mathrm{RL}} = 0$ and $\bar{g}^{(2)}_{\mathrm{RL}} \simeq 1/2\langle \hat{n} \rangle$ in the wings (where $\langle \hat{n}\rangle\!=\!1$ for the $|1,1\rangle $ case), with the resulting maximum visibility of $V=1$. If, on the other hand, the contribution of the $|2,2\rangle$ and higher-$n$ components is not negligible (which is the case, for example, of $\langle \hat{n} \rangle\simeq 0.1$) then the raw coincidence counts at the HOM dip and the respective normalised correlation function no longer equal to zero; in fact, the full SPDC state for arbitrary $\alpha<1$ leads to a HOM dip minimum of $\bar{g}^{(2)}_{\mathrm{RL}} = 1$ and $\bar{g}^{(2)}_{\mathrm{RL}} = 2 + 1/2\langle \hat{n} \rangle$ in the wings, which in turn results in a reduced visibility of $V=1-1/(2+1/2\langle\hat{n}\rangle)$. Despite this reduction, the visibility is still very close to $V=1$ in the weak-gain regime ($\langle \hat{n}\rangle \ll 1 $) where  it scales as $V\simeq 1-2\langle \hat{n}\rangle $.

The process of four-wave mixing of matter-waves gives rise to an output state analogous to the above SPDC state for each pair of correlated modes (see, \textit{e.g.}, \cite{Perrin-08,Ogren-09} and Supplementary Note 1). Indeed, the fraction of atoms converted from the source BEC to all scattering modes is typically less than $5\%$, which justifies the use of the undepleted pump approximation. 
The typical occupation numbers of the scattered modes are, however, beyond the extreme of a very weak gain. In our simulations, the mode occupation on the scattering halo is on the order of $0.1$ and therefore, even in the simplified analytical treatment of the process, the output state of any given pair of correlated modes cannot be approximated by the truncated state $|0,0\rangle + \alpha|1,1\rangle$ or indeed the idealised twin-Fock state $|1,1\rangle$.
\\
\\
\textbf{Scaling with mode population and experimental considerations.} At the basic level, our proposal only relies on the existence of the aforementioned pair-correlations between scattered atoms, with the strength of the correlations affecting the visibility of the HOM dip. For a sufficiently homogeneous source BEC \cite{Ogren-09,Ogren-10}, the correlations 
and thus the visibility $V$ effectively depend only on the average mode population $\langle\hat{n}\rangle$ in the scattering halo, with a scaling of $V$ on $\langle\hat{n}\rangle$ given by 
$V=1-1/(2+1/2\langle\hat{n}\rangle)$ by our analytic model.
Dependence of $\langle\hat{n}\rangle$ on system parameters 
such as the total number of atoms in the initial BEC, trap frequencies, and collision duration is well understood both theoretically and experimentally \cite{Perrin-08,Krachmalnikoff:2010,Jaskula-10,Kheruntsyan:2012}, and each can be sufficiently controlled such that a suitable mode population of $\langle\hat{n}\rangle\lesssim 1$ can, in principle, be targeted. 
There lies, however, a need for optimisation: very small populations are preferred for higher visibility, but they inevitably lead to a low signal-to-noise, hence requiring a potentially very large number of experimental runs for acquiring statistically significant data. Large occupations, on the other hand, lead to higher signal-to-noise, but also to a degradation of the visibility towards the nonclassical threshold of $V=0.5$.
The mode population of $\sim 0.1$ resulting from our numerical simulations appears to be a reasonable compromise; following the scaling of the visibility with $\langle\hat{n}\rangle$ predicted by the simple analytic 
model, it appears that one could safely increase the population to $\sim 0.2$ before a nonclassical threshold is reached to within a typical  uncertainty of $\sim12$\% (as per quoted value of $V\simeq 0.69\pm0.08$) obtained through our simulations.

The proposal is also robust to other experimental considerations such as the implementation of the Bragg pulses; \textit{e.g.}, one may use square Bragg pulses rather than Gaussians. Furthermore, experimental control of the Bragg pulses is 
sufficiently accurate to avoid any degradation of the dip visibility. Modifying the relative timing of the $\pi$ and $\pi/2$ pulses by few percent in our simulations does not explicitly affect the dip visibility, rather only the period of the 
oscillations in the wings of $\overline{g}^{(2)}_{\mathrm{RL}}(t_4)$. This may lead to a systematic change in the calculated dip visibility, however, this is overwhelmed by the uncertainty of $12$\% which accounts for the fluctuations of $\overline{g}^{(2)}_{\mathrm{RL}}(t_4)$ about the mean.

Importantly, we expect that the fundamentally new aspects of the matter-wave setup, namely the multimode nature of the scattering halo and the differences from the archetypical HOM input state of $|1,1\rangle$, as well as the 
specific measurement protocol we have proposed for dealing with these new aspects, are broadly applicable to other related matter-wave setups that generate pair-correlated atoms. These include molecular 
dissociation \cite{Savage:06}, an elongated BEC in a parametrically shaken trap \cite{Vienna-twins}, or degenerate four-wave mixing in an optical lattice \cite{Moelmer:lattice,westbrook:beams}. In the present work, we focus on condensate collisions only due to the accurate characterisation, both experimental and 
theoretical, of the atom-atom correlations, including in a variety of collision geometries \cite{Perrin:2007,Perrin-08,Krachmalnikoff:2010,Jaskula-10,Kheruntsyan:2012}.
\\
\\
\textbf{\large{Discussion}}\\
In summary, we have shown that an atom-optics analogue of the Hong-Ou-Mandel
effect can be realised using colliding condensates and laser induced
Bragg pulses. The HOM dip visibility greater than $50$\% 
implies that the atom-atom correlations in this process cannot be
described by classical stochastic random variables. Generation and
detection of such quantum correlations in matter waves can serve as
precursors to stronger tests of quantum mechanics such as those implied
by a Bell inequality violation and the Einstein-Podolsky-Rosen paradox
\cite{PhysRevA.86.032115}. In particular, the experimental demonstration
of the atom-optics HOM effect would serve as a suitable starting point
to experimentally demonstrate a violation of a Bell inequality using
an atom-optics adaptation of the Rarity-Tapster setup \cite{Rarity:90}.
In this setup, one would tune the Bragg pulses as to realise two separate HOM-interferometer arms, enabling to mix \textit{two}
angle-resolved pairs of momentum modes from the collisional halo, such as ($\mathbf{k},\mathbf{q}$)
and ($-\mathbf{k},-\mathbf{q}$), which would then form the basis of a Bell state
$|\Psi\rangle=\frac{1}{\sqrt{2}}(|\mathbf{k},-\mathbf{k}\rangle+|\mathbf{q},-\mathbf{q}\rangle)$  \cite{Lewis-Swan-KK}. \\
\\
\textbf{Methods} \\
\small{\textbf{Stochastic Bogoliubov approach for simulations.} 
To simulate the collision dynamics, we use the time-dependent stochastic Bogoliubov
approach \cite{Krachmalnikoff:2010,Deuar-11} used previously to accurately model a number of condensate collision experiments 
\cite{Krachmalnikoff:2010,Jaskula-10,Kheruntsyan:2012}. In this approach, 
the atomic field operator is split into $\hat{\psi}\left(\mathbf{r},t\right)\!=\!\psi\left(\mathbf{r},t\right)+\hat{\delta}\left(\mathbf{r},t\right)$,
where $\psi$ is the mean-field component describing the 
source condensates and $\hat{\delta}$ is the fluctuating component
(treated to lowest order in perturbation theory) describing the scattered
atoms. The mean-field component evolves according to the standard time-dependent
Gross-Pitaevskii (GP) equation, where the initial state is taken in the form of 
$\psi\left(\mathbf{r},0\right)=\sqrt{\rho_{0}\left(\mathbf{r}\right)/2}\left(e^{ik_{0}z}+e^{-ik_{0}z}\right)$.
This models an instantaneous splitting at $t=0$ of a zero-temperature
condensate in a coherent state into two halves which subsequently
evolve in free space, where $\rho_{0}\left(\mathbf{r}\right)$ is
the particle number density of the initial (trapped) sample before
splitting.

The fluctuating component is simulated using the stochastic counterpart of the 
Heisenberg operator equations of motion \cite{Ogren-09,Krachmalnikoff:2010}, $i\hbar\partial_t\hat{\delta}(\mathbf{r},t)= \mathcal{H}_{0}(\mathbf{r},t)\hat{\delta}+\Upsilon(\mathbf{r},t)\hat{\delta}^{\dagger}$,
in the positive $P$-representation with the vacuum initial state. 
Here $\mathcal{H}_{0}\left(\mathbf{r},t\right)\!=\!-\frac{\hbar^{2}}{2m}\nabla^{2}\!+\!2U|\psi\left(\mathbf{r},t\right)|^{2}+V_{\mathrm{BP}}(\mathbf{r},t)$
represents the kinetic energy term, an effective mean-field potential, plus the lattice potential $V_{\mathrm{BP}}(\mathbf{r},t)$ imposed by the Bragg pulses,
whereas $\Upsilon\left(\mathbf{r},t\right)=U\psi\left(\mathbf{r},t\right)^{2}$
is an effective coupling responsible for the spontaneous pair-production
of scattered atoms. The interaction constant $U$ is given by $U=4\pi\hbar^{2}a/m$, where
 $m$ is the atomic mass, and \emph{$a$} is the \emph{s}-wave scattering length.
\\
\\
\small{\textbf{Details of Bragg pulses.}
The Bragg pulses 
are realised by two interfering laser beams (assumed
for simplicity to have a uniform intensity across the atomic
cloud and zero relative phase) that create a periodic lattice potential
$V_{\mathrm{BP}}\left(\mathbf{r},t\right)\!=\!\frac{1}{2}V_{L}(t)\mathrm{cos}\left( 2\mathbf{k}_{L}\cdot\mathbf{r} \right)$,
where $V_{L}(t)$ is the 
lattice depth and $\mathbf{k}_{L}\!=\!\frac{1}{2}(\mathbf{k}_{L,2}\!-\!\mathbf{k}_{L,1})$
is the lattice vector determined by the wave-vectors $\mathbf{k}_{L,i}$ ($i\!=\!1,2$) of the two lasers, and tuned to $|\mathbf{k}_{L}|\!=\!k_r$.
The Bragg pulses couple momentum modes $\mathbf{k}_{i}$ and $\mathbf{k}_j\!=\! \mathbf{k}_i \!-\! 2\mathbf{k}_L$, 
satisfying momentum and energy conservation (up to a finite width due
to energy-time uncertainty \cite{Batelaan-00}).
The lattice depth is ramped up (down) according to 
$V_{L}(t) \!=\! V_0\mathrm{exp}[-(t-t_2)^2/2\tau_{\pi}^2] + \frac{1}{2}V_0\mathrm{exp}[-(t-t_3)^2/2\tau_{\pi/2}^2]$,
where $t_{2(3)}$ is the pulse centre, while $\tau_{\pi(\pi/2)}$ is the 
pulse duration which governs the transfer of atomic population between the targeted momentum modes: a $\pi$-pulse is defined by $\tau_{\pi}\!=\!\sqrt{2\pi}\hbar/{V}_0$ and converts the entire population
from one momentum mode to the other, while a $\pi/2$-pulse is defined
by $\tau_{\pi/2}\!=\!\sqrt{\pi}\hbar/(\sqrt{2}{V}_0)$ and converts only
half of the population.
\\
\\
\small{\textbf{Aspects of measurement after expansion.}
In practice, the atom-atom correlations quantifying the HOM interference are measured in position space after the low-density scattering halo expands ballistically in free space and falls under gravity onto an atom detector. The detector records the arrival times and positions of individual atoms, which is literally the case for metastable helium atoms considered here \cite{Perrin:2007,Krachmalnikoff:2010,Jaskula-10,Kheruntsyan:2012,metastable-RMP}. The arrival times and positions are used to reconstruct the three-dimensional velocity (momentum) distribution before expansion, as well as the atom-atom coincidences for any desired pair of momentum vectors. In our simulations and the proposed geometry of the experiment, the entire system (including the Bragg pulses) maintains reflectional symmetry about the $yz$-plane, with $z$ being the vertical direction. Therefore the effect of gravity can be completely ignored as it does not introduce any asymmetry to the momentum distribution of the atoms and their correlations on the equatorial plane of the halo or indeed any other plane parallel to it.

\bibliographystyle{naturemag}

\begin{thebibliography}{10}
\expandafter\ifx\csname url\endcsname\relax
  \def\url#1{\texttt{#1}}\fi
\expandafter\ifx\csname urlprefix\endcsname\relax\def\urlprefix{URL }\fi
\providecommand{\bibinfo}[2]{#2}
\providecommand{\eprint}[2][]{\url{#2}}

\bibitem{HOM-87}
\bibinfo{author}{Hong, C.~K.}, \bibinfo{author}{Ou, Z.~Y.} \&
  \bibinfo{author}{Mandel, L.}
\newblock \bibinfo{title}{Measurement of subpicosecond time intervals between
  two photons by interference}.
\newblock \emph{\bibinfo{journal}{Phys. Rev. Lett.}}
  \textbf{\bibinfo{volume}{59}}, \bibinfo{pages}{2044--2046}
  (\bibinfo{year}{1987}).

\bibitem{Ou-Mandel-88}
\bibinfo{author}{Ou, Z.~Y.} \& \bibinfo{author}{Mandel, L.}
\newblock \bibinfo{title}{Violation of {B}ell's inequality and classical
  probability in a two-photon correlation experiment}.
\newblock \emph{\bibinfo{journal}{Phys. Rev. Lett.}}
  \textbf{\bibinfo{volume}{61}}, \bibinfo{pages}{50--53}
  (\bibinfo{year}{1988}).

\bibitem{Linear_quantum_computation}
\bibinfo{author}{Knill, E.}, \bibinfo{author}{Laflamme, R.} \&
  \bibinfo{author}{Milburn, G.~J.}
\newblock \bibinfo{title}{A scheme for efficient quantum computation with
  linear optics}.
\newblock \emph{\bibinfo{journal}{{Nature}}} \textbf{\bibinfo{volume}{409}},
  \bibinfo{pages}{46} (\bibinfo{year}{2001}).

\bibitem{Dowling-NOON}
\bibinfo{author}{Kok, P.}, \bibinfo{author}{Lee, H.} \&
  \bibinfo{author}{Dowling, J.~P.}
\newblock \bibinfo{title}{Creation of large-photon-number path entanglement
  conditioned on photodetection}.
\newblock \emph{\bibinfo{journal}{Phys. Rev. A}} \textbf{\bibinfo{volume}{65}},
  \bibinfo{pages}{052104} (\bibinfo{year}{2002}).

\bibitem{Duan-Monroe-2010}
\bibinfo{author}{Duan, L.-M.} \& \bibinfo{author}{Monroe, C.}
\newblock \bibinfo{title}{\textit{Colloquium}: Quantum networks with trapped
  ions}.
\newblock \emph{\bibinfo{journal}{Rev. Mod. Phys.}}
  \textbf{\bibinfo{volume}{82}}, \bibinfo{pages}{1209--1224}
  (\bibinfo{year}{2010}).

\bibitem{HOM-microwave}
\bibinfo{author}{Lang, C.} \emph{et~al.}
\newblock \bibinfo{title}{Correlations, indistinguishability and entanglement
  in {H}ong-{O}u-{M}andel experiments at microwave frequencies}.
\newblock \emph{\bibinfo{journal}{Nature Phys.}} \textbf{\bibinfo{volume}{9}},
  \bibinfo{pages}{345--348} (\bibinfo{year}{2013}).

\bibitem{Jaskula-10}
\bibinfo{author}{Jaskula, J.} \emph{et~al.}
\newblock \bibinfo{title}{Sub-{P}oissonian number differences in four-wave
  mixing of matter waves}.
\newblock \emph{\bibinfo{journal}{Phys. Rev. Lett.}}
  \textbf{\bibinfo{volume}{105}}, \bibinfo{pages}{190402}
  (\bibinfo{year}{2010}).

\bibitem{Krachmalnikoff:2010}
\bibinfo{author}{Krachmalnicoff, V.} \emph{et~al.}
\newblock \bibinfo{title}{Spontaneous four-wave mixing of de {B}roglie waves:
  Beyond optics}.
\newblock \emph{\bibinfo{journal}{Phys. Rev. Lett.}}
  \textbf{\bibinfo{volume}{104}}, \bibinfo{pages}{150402}
  (\bibinfo{year}{2010}).

\bibitem{Perrin:2007}
\bibinfo{author}{Perrin, A.} \emph{et~al.}
\newblock \bibinfo{title}{Observation of atom pairs in spontaneous four-wave
  mixing of two colliding {B}ose-{E}instein condensates}.
\newblock \emph{\bibinfo{journal}{Phys. Rev. Lett.}}
  \textbf{\bibinfo{volume}{99}}, \bibinfo{pages}{150405}
  (\bibinfo{year}{2007}).

\bibitem{Perrin-08}
\bibinfo{author}{Perrin, A.} \emph{et~al.}
\newblock \bibinfo{title}{Atomic four-wave mixing via condensate collisions}.
\newblock \emph{\bibinfo{journal}{New J. Phys.}} \textbf{\bibinfo{volume}{10}},
  \bibinfo{pages}{045021} (\bibinfo{year}{2008}).

\bibitem{Kheruntsyan:2012}
\bibinfo{author}{Kheruntsyan, K.~V.} \emph{et~al.}
\newblock \bibinfo{title}{Violation of the {C}auchy-{S}chwarz inequality with
  matter waves}.
\newblock \emph{\bibinfo{journal}{Phys. Rev. Lett.}}
  \textbf{\bibinfo{volume}{108}}, \bibinfo{pages}{260401}
  (\bibinfo{year}{2012}).

\bibitem{Kozuma-99}
\bibinfo{author}{Kozuma, M.} \emph{et~al.}
\newblock \bibinfo{title}{Coherent splitting of {B}ose-{E}instein condensed
  atoms with optically induced {B}ragg diffraction}.
\newblock \emph{\bibinfo{journal}{Phys. Rev. Lett.}}
  \textbf{\bibinfo{volume}{82}}, \bibinfo{pages}{871--875}
  (\bibinfo{year}{1999}).

\bibitem{meystre-atom-optics}
\bibinfo{author}{Meystre, P.}
\newblock \emph{\bibinfo{title}{Atom Optics}}
  (\bibinfo{publisher}{Springer-Verlag}, \bibinfo{address}{New York},
  \bibinfo{year}{2001}).

\bibitem{Vienna-twins}
\bibinfo{author}{B\"ucker, R.} \emph{et~al.}
\newblock \bibinfo{title}{Twin-atom beams}.
\newblock \emph{\bibinfo{journal}{Nature Phys.}} \textbf{\bibinfo{volume}{7}},
  \bibinfo{pages}{608} (\bibinfo{year}{2011}).

\bibitem{Hannover-twins}
\bibinfo{author}{L\"ucke, B.} \emph{et~al.}
\newblock \bibinfo{title}{Twin matter waves for interferometry beyond the
  classical limit}.
\newblock \emph{\bibinfo{journal}{Science}} \textbf{\bibinfo{volume}{334}},
  \bibinfo{pages}{773} (\bibinfo{year}{2011}).

\bibitem{Oberthaler-quadratures}
\bibinfo{author}{Gross, C.} \emph{et~al.}
\newblock \bibinfo{title}{Atomic homodyne detection of continuous-variable
  entangled twin-atom states}.
\newblock \emph{\bibinfo{journal}{{Nature}}} \textbf{\bibinfo{volume}{480}},
  \bibinfo{pages}{219} (\bibinfo{year}{2011}).

\bibitem{Rarity:90}
\bibinfo{author}{Rarity, J.~G.} \& \bibinfo{author}{Tapster, P.~R.}
\newblock \bibinfo{title}{Experimental violation of {B}ell's inequality based
  on phase and momentum}.
\newblock \emph{\bibinfo{journal}{Phys. Rev. Lett.}}
  \textbf{\bibinfo{volume}{64}}, \bibinfo{pages}{2495--2498}
  (\bibinfo{year}{1990}).

\bibitem{Gardiner:06}
\bibinfo{author}{Norrie, A.~A.}, \bibinfo{author}{Ballagh, R.~J.} \&
  \bibinfo{author}{Gardiner, C.~W.}
\newblock \bibinfo{title}{Quantum turbulence and correlations in
  {B}ose-{E}instein condensate collisions}.
\newblock \emph{\bibinfo{journal}{Phys. Rev. A}} \textbf{\bibinfo{volume}{73}},
  \bibinfo{pages}{043617} (\bibinfo{year}{2006}).

\bibitem{Savage:06}
\bibinfo{author}{Savage, C.~M.}, \bibinfo{author}{Schwenn, P.~E.} \&
  \bibinfo{author}{Kheruntsyan, K.~V.}
\newblock \bibinfo{title}{First-principles quantum simulations of dissociation
  of molecular condensates: {A}tom correlations in momentum space}.
\newblock \emph{\bibinfo{journal}{Phys. Rev. A}} \textbf{\bibinfo{volume}{74}},
  \bibinfo{pages}{033620} (\bibinfo{year}{2006}).

\bibitem{Deuar:07}
\bibinfo{author}{Deuar, P.} \& \bibinfo{author}{Drummond, P.~D.}
\newblock \bibinfo{title}{Correlations in a {BEC} collision: First-principles
  quantum dynamics with 150\,000 atoms}.
\newblock \emph{\bibinfo{journal}{Phys. Rev. Lett.}}
  \textbf{\bibinfo{volume}{98}}, \bibinfo{pages}{120402}
  (\bibinfo{year}{2007}).

\bibitem{Moelmer:2008}
\bibinfo{author}{M\o{}lmer, K.} \emph{et~al.}
\newblock \bibinfo{title}{Hanbury {B}rown and {T}wiss correlations in atoms
  scattered from colliding condensates}.
\newblock \emph{\bibinfo{journal}{Phys. Rev. A}} \textbf{\bibinfo{volume}{77}},
  \bibinfo{pages}{033601} (\bibinfo{year}{2008}).

\bibitem{Ogren-09}
\bibinfo{author}{Ogren, M.} \& \bibinfo{author}{Kheruntsyan, K.~V.}
\newblock \bibinfo{title}{Atom-atom correlations in colliding {B}ose-{E}instein
  condensates}.
\newblock \emph{\bibinfo{journal}{Phys. Rev. A.}}
  \textbf{\bibinfo{volume}{79}}, \bibinfo{pages}{021606}
  (\bibinfo{year}{2009}).

\bibitem{Deuar-11}
\bibinfo{author}{Deuar, P.}, \bibinfo{author}{Chwede\'{n}czuk, J.},
  \bibinfo{author}{Trippenbach, M.} \& \bibinfo{author}{Zi\'{n}, P.}
\newblock \bibinfo{title}{Bogoliubov dynamics of condensate collisions using
  the positive-{$P$} representation}.
\newblock \emph{\bibinfo{journal}{Phys. Rev. A}} \textbf{\bibinfo{volume}{83}},
  \bibinfo{pages}{063625} (\bibinfo{year}{2011}).

\bibitem{Batelaan-00}
\bibinfo{author}{Batelaan, H.}
\newblock \bibinfo{title}{The {K}apitza-{D}irac effect}.
\newblock \emph{\bibinfo{journal}{Contemp. Phys.}}
  \textbf{\bibinfo{volume}{41}}, \bibinfo{pages}{369--381}
  (\bibinfo{year}{2000}).

\bibitem{Rarity-89}
\bibinfo{author}{Rarity, J.~G.} \& \bibinfo{author}{Tapster, P.~R.}
\newblock \bibinfo{title}{Two-color photons and nonlocality in fourth-order
  interference}.
\newblock \emph{\bibinfo{journal}{Phys. Rev. A}} \textbf{\bibinfo{volume}{41}},
  \bibinfo{pages}{5139--5146} (\bibinfo{year}{1990}).

\bibitem{Braunstein-review}
\bibinfo{author}{Braunstein, S.~L.} \& \bibinfo{author}{van Loock, P.}
\newblock \bibinfo{title}{Quantum information with continuous variables}.
\newblock \emph{\bibinfo{journal}{Rev. Mod. Phys.}}
  \textbf{\bibinfo{volume}{77}}, \bibinfo{pages}{513--577}
  (\bibinfo{year}{2005}).

\bibitem{Ogren-10}
\bibinfo{author}{\"Ogren, M.} \& \bibinfo{author}{Kheruntsyan, K.~V.}
\newblock \bibinfo{title}{Role of spatial inhomogeneity in dissociation of
  trapped molecular condensates}.
\newblock \emph{\bibinfo{journal}{Phys. Rev. A}} \textbf{\bibinfo{volume}{82}},
  \bibinfo{pages}{013641} (\bibinfo{year}{2010}).

\bibitem{Moelmer:lattice}
\bibinfo{author}{Hilligs\o{}e, K.~M.} \& \bibinfo{author}{M\o{}lmer, K.}
\newblock \bibinfo{title}{Phase-matched four wave mixing and quantum beam
  splitting of matter waves in a periodic potential}.
\newblock \emph{\bibinfo{journal}{Phys. Rev. A}} \textbf{\bibinfo{volume}{71}},
  \bibinfo{pages}{041602} (\bibinfo{year}{2005}).

\bibitem{westbrook:beams}
\bibinfo{author}{Bonneau, M.} \emph{et~al.}
\newblock \bibinfo{title}{Tunable source of correlated atom beams}.
\newblock \emph{\bibinfo{journal}{Phys. Rev. A}} \textbf{\bibinfo{volume}{87}},
  \bibinfo{pages}{061603} (\bibinfo{year}{2013}).

\bibitem{PhysRevA.86.032115}
\bibinfo{author}{Kofler, J.} \emph{et~al.}
\newblock \bibinfo{title}{Einstein-{P}odolsky-{R}osen correlations from
  colliding {B}ose-{E}instein condensates}.
\newblock \emph{\bibinfo{journal}{Phys. Rev. A}} \textbf{\bibinfo{volume}{86}},
  \bibinfo{pages}{032115} (\bibinfo{year}{2012}).

\bibitem{Lewis-Swan-KK}
\bibinfo{note}{R. J. Lewis-Swan and K. V. Kheruntsyan, Book of Abstracts of
  ICAP 2012 -- The 23rd International Conference on Atomic Physics (23-27 July
  2012, Ecole Polytechnique, Palaiseau, France).}

\bibitem{metastable-RMP}
\bibinfo{author}{Vassen, W.} \emph{et~al.}
\newblock \bibinfo{title}{Cold and trapped metastable noble gases}.
\newblock \emph{\bibinfo{journal}{Rev. Mod. Phys.}}
  \textbf{\bibinfo{volume}{84}}, \bibinfo{pages}{175--210}
  (\bibinfo{year}{2012}).

\end{thebibliography}

~\\
\\
\textbf{\large{Acknowledgements}}\\
\small{The authors acknowledge discussions with C. I. Westbrook and D. Boiron.
K.V.K. acknowledges support by the ARC Future Fellowship award FT100100285.}
\\
\\
\textbf{\large{Author contributions}}\\
\small{R.J.L-S. conducted the numerical simulations and derived the analytic treatment. Both authors contributed extensively to the conceptual formulation of the physics, the interpretation of the data, and writing the manuscript. }
\\
\\
\textbf{\large{Additional information}}\\ 
\small{\textbf{Supplementary Information} accompanies this paper at http://www.nature.com/
naturecommunications.}\\
\\
\small{\textbf{Competing financial interests:} The authors declare no competing financial interests.}

\end{document}